\documentstyle[12pt]{article}
\title{Some eigenstates for a model associated with solutions
       of tetrahedron equation}
\author{I.G.~Korepanov\\
\footnotesize $\matrix{
       \hbox{Chelyabinsk University of Technology}\cr
       \hbox{76 Lenin av., Chelyabinsk 454080, Russia}
       }$}
\date{January 1997}
\def\be{\begin{equation}}
\def\ee{\end{equation}}
\makeatletter
\long\def\@makecaption#1#2{\vskip 10\p@ \hbox to\hsize{\hfil#1\hfil}}
\makeatother
\begin{document}
\maketitle

\begin{abstract}
Here we present some eigenstates for a $2+1$-dimensional model associated
with a solution of the tetrahedron equation.
The eigenstates include those ``particle-like'' (namely one-particle
and two-particle ones), constructed in analogy
with the usual $1+1$-dimensional Bethe ansatz, and some simple
``string-like'' ones.
\end{abstract}

\section*{Introduction}

The problem of extending the Bethe ansatz onto the $2+1$-dimensional models,
or maybe finding some other method for constructing their eigenvectors,
is surely of great importance. In this paper, we are going to present
some results obtained in this direction.

Our model will be a model on the cubic lattice: the lattice
vertices are points whose all three coordinates are integers.
The lattice will be assumed infinite in all directions,
unless the contrary is stated explicitly. Thus, the calculations
in this paper will be in part formally-algebraic.
In each vertex there is an ``$S$-operator'' acting in the tensor
product of three linear spaces attached to the {\em links},
as in paper~\cite{k2}. To be exact, it will be convenient for us
to assume that there is a transposed matrix $S^{\rm T}$ in each
vertex. This is because we will need some objects called
``vacuum vectors'', and the ``vacuum covectors'' of $S$ studied
in~\cite{k2} are ``vacuum vectors'' of $S^{\rm T}$.

The transfer matrix we will deal with will be a ``diagonal'' one:
it is cut out of the lattice by two planes perpendicular to the vector
$(1,1,1)$ in such a way that it consist of separate, not linked to
each other, ``anti-tank hedgehogs'' (vertices). In each of those
planes, the intersection with the cubic lattice yields
a kagome lattice consisting, as known, of triangles and hexagons.
We can group all vertices of the kagome lattice in triples---vertices
of triangles---in such a way that
the transfer matrix acts on each triangle separately, turning it
inside out and making a linear transformation in the tensor product
of three corresponding subspaces.

The mentioned product of three subspaces is comprised of the
0-, 1-, 2- and 3-particle sectors. According to
papers~\cite{k1,k2}, the sectors with even and odd particle numbers
do not mix together and, moreover, in the even sector the $S$-operator
acts as an identical unity.
The 1- and 3-particle sectors do mix together, but it turns out
that {\em there are two eigenvectors of the $S$-operator in the
one-particle sector\/}, with eigenvalues $1$ and $-1$.
Their explicit form can be extracted out of the end
of p.~94 and the beginning of p.~95 of~\cite{k2}.
Namely, denote as $(x,y,z)^{\rm T}$ a one-particle state describing
the situation when the ``amplitudes'' for a particle to be in the
1st, 2nd and 3rd spaces are $x$, $y$ and $z$. According to~\cite{k2},
and taking into account the fact that we are considering the vectors
without a 3-particle component, we can take any ``isotropic'' vector,
i.e.\ such that
$x^2-y^2+z^2=0$, as the eigenvector corresponding to the eigenvalue $1$,
and any other ``isotropic'' vector similarly for the eigenvalue $-1$.
Let us write those vectors as
\be
(\sin\lambda_+,1,\cos\lambda_+)^{\rm T}
\label{1}
\ee
and
\be
(\sin\lambda_-,1,\cos\lambda_-)^{\rm T},
\label{2}
\ee
respectively.

A vector $(x,y,z)^{\rm T}$ is a linear combination of the vectors
(\ref{1}) and (\ref{2}) iff
\be
y=ax+bz,
\label{3}
\ee
where
\be
a={\cos\lambda_- -\cos\lambda_+ \over \sin(\lambda_+ -\lambda_-)},
\qquad
b={\sin\lambda_+ -\sin\lambda_- \over \sin(\lambda_+ -\lambda_-)}.
\label{3.5}
\ee

\section{One-particle states}
\label{sec-onep}

Consider now the whole kagome lattice. For it, the one-particle space
is the direct sum of one-particle spaces over all its vertices
multiplied by vacuums in other places. To indicate a one-particle
vector $\varphi$ means to attach a number---amplitude $\varphi_A$---to
each vertex $A$ of the kagome lattice.
It is evident that in order to ensure that the vector never comes out
of the one-particle space on applying to it any number of transfer matrices,
we must take it to be an {\em eigenvector\/} of the composition
of a transfer matrix and a proper lattice shift. The latter arises
 from the fact that a transfer matrix, when acting on a ``slice vector'',
turns inside out half of the triangles of the kagome lattice to which
that vector belongs, thus moving the lines.

So, let us write down the conditions for a vector in the one-particle space
to be an eigenvector. Consider the following picture (Fig.~\ref{fig1})
\begin{figure}
\begin{center}
\unitlength=1.00mm
\linethickness{0.4pt}
\begin{picture}(50.00,50.00)
\put(10.00,0.00){\line(3,5){30.00}}
\put(40.00,0.00){\line(-3,5){30.00}}
\put(0.00,45.00){\line(1,0){50.00}}
\put(50.00,5.00){\line(-1,0){50.00}}
\put(8.00,47.00){\makebox(0,0)[rb]{$A$}}
\put(42.00,47.00){\makebox(0,0)[lb]{$B$}}
\put(42.00,3.00){\makebox(0,0)[lt]{$E$}}
\put(8.00,3.00){\makebox(0,0)[rt]{$D$}}
\put(29.00,25.00){\makebox(0,0)[lc]{$C$}}
\put(25.00,22.00){\vector(0,-1){5.00}}
\put(13.00,5.00){\circle*{1.00}}
\put(13.00,45.00){\circle*{1.00}}
\put(37.00,45.00){\circle*{1.00}}
\put(37.00,5.00){\circle*{1.00}}
\put(25.00,25.00){\circle*{1.00}}
\put(34.00,7.00){\vector(-4,3){5.00}}
\put(16.00,7.00){\vector(4,3){5.00}}
\end{picture}
\end{center}
\caption{}
\label{fig1}
\end{figure}
representing a fragment of the kagome lattice.
Here the triangle $DCE$ is going to be turned inside out, while
the triangle $BCA$ has been obtained by turning inside out a triangle
on the previous step. So, the two conditions arise:
\be
\varphi_C=a\varphi_D+b\varphi_E
\label{4}
\ee
and
\be
\varphi_C=a\varphi_B+b\varphi_A
\label{5}
\ee
(compare~(\ref{3})).

When the triangle $DCE$ is turned inside out, it yields
a triangle $D'C'E'$ (Fig.~\ref{fig2}),
\begin{figure}
\begin{center}
\unitlength=1.00mm
\linethickness{0.4pt}
\begin{picture}(50.00,30.00)
\put(0.00,25.00){\line(1,0){50.00}}
\put(8.00,27.00){\makebox(0,0)[rb]{$E'$}}
\put(42.00,27.00){\makebox(0,0)[lb]{$D'$}}
\put(29.00,5.00){\makebox(0,0)[lc]{$C'$}}
\put(25.00,14.00){\vector(0,-1){5.00}}
\put(22.00,0.00){\line(3,5){18.00}}
\put(28.00,0.00){\line(-3,5){18.00}}
\put(25.00,5.00){\circle*{1.00}}
\put(37.00,25.00){\circle*{1.00}}
\put(13.00,25.00){\circle*{1.00}}
\put(22.00,19.00){\vector(-4,3){5.00}}
\put(28.00,19.00){\vector(4,3){5.00}}
\end{picture}
\end{center}
\caption{}
\label{fig2}
\end{figure}
and the new ``field'' variables
are expressed through the old ones as
\be
\pmatrix{\varphi_{D'}\cr \varphi_{E'}}=
\pmatrix{\alpha & \beta \cr \gamma & \delta}
\pmatrix{\varphi_D\cr \varphi_E},
\label{6}
\ee
where
$$
\pmatrix{\alpha & \beta \cr \gamma & \delta}=
{1\over \sin(\lambda_+ -\lambda_-)}
\pmatrix{\sin(\lambda_+ +\lambda_-) & -2\sin\lambda_+ \sin\lambda_- \cr
2\cos\lambda_+ \cos\lambda_-        & -\sin(\lambda_+ +\lambda_-)}.
$$
Note that
\be
\alpha=-\delta,\qquad \alpha^2+\beta\gamma=1.
\label{6.5}
\ee

The points $E'$ and $D'$ of the ``new'' lattice are analogs of
the points $A$ and $B$ correspondingly belonging to the ``old'' lattice.
Thus, in order to obtain on the new lattice a vector proportional
to the vector on the old lattice, we must require that
\be
{\varphi_{E'}\over \varphi_{D'}}={\varphi_A\over \varphi_B}.
\label{7}
\ee
If this condition holds, one can extend
both the old vector $\varphi$ and the new ``primed'' vector periodically
onto the whole lattice and in such way that the new one will
be proportional to the (shifted) old one.

The condition~(\ref{7}) together with (\ref{4}, \ref{5}, \ref{6}) is enough
to obtain $\varphi_A$ and $\varphi_B$ (as well as $\varphi_C$,
$\varphi_{D'}$ and $\varphi_{E'}$) out of given $\varphi_D$ and
$\varphi_E$. Thus, only one essential free parameter, e.g.\ 
$\varphi_D / \varphi_E$, remains
for our construction of one-particle eigenvectors.

\section{Two-particle states}
\label{sec-twop}

How can the superposition of two one-particle states of
Section~\ref{sec-onep} look like? The experience of studying
the $2+1$-dimensional {\em classical\/} integrable models hints that
probably the ``scattering'' of two particles on one another
must be trivial, i.e.\ it makes sense to assume
for the ``amplitude of the event that two particles are in two different
points $F$ and $G$ of the lattice'' the form
\be
\Phi_{FG}=\varphi_F \psi_G + \varphi_G \psi_F,
\label{8}
\ee
where $\varphi_{\ldots}$ and $\psi_{\ldots}$ are one-particle amplitudes
like those constructed in Section~\ref{sec-onep}.

To see how $\Phi_{FG}$ transforms under the action of transfer matrix,
let us decompose $\varphi_{\ldots}$ and $\psi_{\ldots}$, considered
as functions of $F$, in sums
over triangles of the type $DCE$ in Fig.~\ref{fig1}, i.e.\
represent $\varphi_F$ and $\psi_F$ as sums of summands each of which
equals zero if $F$ lies beyond the corresponding triangle.
In this way, $\Phi_{FG}$ naturally decomposes in a sum over (non-ordered)
pairs of such triangles, including pairs of two coinciding triangles.
We want that $\Phi_{FG}$ be transformed by the transfer matrix in an
expression of the same form~(\ref{8}), with $\varphi_{\ldots}$ and
$\psi_{\ldots}$ changed to their images with respect to this action.

It is easy to see that this holds automatically if $F$ and $G$ belong
to {\em different\/} triangles. So, it remains to consider the case
where $F$ and $G$  belong to the same triangle, say triangle~$DCE$
in Fig.~\ref{fig1}. When this triangle is transformed by the transfer
matrix in the triangle~$D'C'E'$ of Fig.~\ref{fig2}, the one-particle
amplitudes are transformed according to~(\ref{6}):
\be
\pmatrix{\varphi_{D'}\cr \varphi_{E'}}=
\pmatrix{\alpha & \beta \cr \gamma & \delta}
\pmatrix{\varphi_D\cr \varphi_E},
\qquad
\pmatrix{\psi_{D'}\cr \psi_{E'}}=
\pmatrix{\alpha & \beta \cr \gamma & \delta}
\pmatrix{\psi_D\cr \psi_E},
\label{9}
\ee
where one must add the conditions of type~(\ref{4}):
\be
\matrix{
\varphi_C=a\varphi_D+b\varphi_E, &
\null \qquad \null &
\varphi_{C'}=a\varphi_{D'}+b\varphi_{E'}, \cr
\psi_C=a\psi_D+b\psi_E, &&
\psi_{C'}=a\psi_{D'}+b\psi_{E'}. }
\label{10}
\ee
On the other hand, the $S$-matrix of the work~\cite{k2} acts trivially,
i.e.\ as a unity matrix, in the 2-particle sector. Thus, it must be
\be
\Phi_{C'D'}=\Phi_{CD},\qquad
\Phi_{C'E'}=\Phi_{CE},\qquad
\Phi_{D'E'}=\Phi_{DE}.
\label{11}
\ee
Together, the formulae (\ref{10}) and (\ref{11}) lead to the following
conditions on the one-particle amplitudes
$\varphi_{\ldots}$ and $\psi_{\ldots}$:
\begin{eqnarray}
\varphi_{E'}\psi_{D'}+\varphi_{D'}\psi_{E'}&=&
\varphi_E\psi_D+\varphi_D\psi_E,\label{12}\\
\varphi_{D'}\psi_{D'}&=&\varphi_D\psi_D,\\
\varphi_{E'}\psi_{E'}&=&\varphi_E\psi_E \label{14}
\end{eqnarray}
(with making no use of the explicit form~(\ref{3.5}) of
coefficients $a$ and $b$).

The three conditions (\ref{12}--\ref{14}) together with~(\ref{9})
and~(\ref{6.5}) give, remarkably, just {\em one\/} condition
\be
-\gamma\varphi_D\psi_D + \alpha(\varphi_D\psi_E+\varphi_E\psi_D) +
\beta\varphi_E\psi_E = 0
\label{15}
\ee
on $\varphi_{\ldots}$ and $\psi_{\ldots}$. Recall that, according
to Section~\ref{sec-onep}, each of the vectors $\varphi$ and $\psi$
is parametrized by one parameter (besides a trivial scalar factor).
Together they are parametrized by two parameters, but
condition~(\ref{15}) subtracts
one parameter. Thus, the two-particle eigenstates constructed
in this section depend on one significant parameter.

\section{The simplest string-like states}

The matrix $S^{\rm T}$, the transposed to the matrix~$S$
of the work~\cite{k2}, has two families of vacuum vectors
(which are vacuum covectors for~$S$, see \cite{k2}, formulae (2.13--2.15),
and also (2.21, 2.22) and the text in a neighborhood of these latter ones).
Here we will restrict ourselves to considering the first family,
i.e.\ the vacuum vectors transformed by $S^{\rm T}$ into themselves:
\be
S^{\rm T} (X^{\rm T}(\zeta) \otimes Y^{\rm T}(\zeta) \otimes
 Z^{\rm T}(\zeta) =
X^{\rm T}(\zeta) \otimes Y^{\rm T}(\zeta) \otimes Z^{\rm T}(\zeta),
\label{16}
\ee
$\zeta$ being a parameter taking values in an elliptic curve
(compare with formula~(1.12) from~\cite{k2}). What we are going to do
in this section can be done with the same success for the second family as
well. Instead of doing that, let us note that the eigenstates from
Sections \ref{sec-onep} and \ref{sec-twop} are built using {\em both\/}
families of vacuum vectors, and in that sense the considerations
in this section are (still) more trivial.

The simplest eigenvectors $\Omega(\zeta)$ of the transfer matrix,
with the eigenvalue 1, are built as follows: fix $\zeta$ and
put in correspondence to each point of type~$D$ (Fig.~\ref{fig1})
of the kagome lattice the vector $X(\zeta)^{\rm T}$, to each point of
type~$D$---the vector $Y(\zeta)^{\rm T}$, and of type~$E$---the
vector~$Z(\zeta)^{\rm T}$.
Then take the tensor product of all those vectors. The formula~(\ref{16})
shows at once that this is indeed an eigenvector with eigenvalue~1.

A little bit more intricate eigenvectors, for which the eigenvalues
in case of a {\em finite\/} lattice are roots of unity, can be
constructed as follows. It is seen from formulae (2.13--2.15)
of paper~\cite{k2}, where enter the values $x$, $y$ and $z$---ratios
of two coordinates of vectors $X$, $Y$ and $Z$ respectively,---that
the triple $X,Y,Z$ will remain vacuum if one makes one of the following
changes:
\begin{eqnarray}
(x,y,z) & \to & (x,{1\over y},{1\over z}), \label{17} \\
(x,y,z) & \to & ({1\over x},y,-{1\over z}), \label{18} \\
(x,y,z) & \to & (-{1\over x},-{1\over y},z). \label{19}
\end{eqnarray}
It can be said that the changes (\ref{17}), (\ref{18}) and (\ref{19})
affect respectively the sides $CE$, $DE$ and $DC$ of the triangle~$DCE$
in Fig.~\ref{fig1}. Obviously, two such changes, if applied successively,
commute with one another.

To construct a vector whose transformation under the action of transfer
matrix is easy to trace, let us act like this: first, select arbitrarily
some strait lines---{\em strings\/}---going along the edges
of the kagome lattice.
Then, take the vector $\Omega(\zeta)$ and change it as follows:
make in each triangle of the type~$DCE$ the transformation(s) of
type (\ref{17}--\ref{19}) if its corresponding side lies in
a selected line.

The obtained vector---let us call it $\Theta$---goes under the action
of transfer matrix $T$ into a vector of a similar form, but with the
properly shifted lines (the latter, let us recall, result from
the intersection of the cubic lattice faces with a plane perpendicular
to the vector $(1,1,1)$, and move in that plane when the plane itself
moves). An eigenvector of $T$ can be now built in the form
\be
\cdots+\omega^{-1}T^{-1}\Theta+\Theta+\omega T\Theta+\omega^2 T^2 \Theta
+\cdots,
\ee
where in the case of a finite lattice the sum must be finite,
and the number $\omega$---a root of unity of a proper degree,
determined by the sizes of the lattice.

\section{Discussion}

In this paper, we are not trying to discuss the ``physical''
consequences of the eigenvectors constructed. Its only modest aim
is to show that the classical Bethe ansatz could be relevant
for $2+1$-dimensional models and that, probably, it makes sense
to search for eigenvectors of another, ``string-like'', form
as well (certainly, it would be of great interest to find
a ``string-like'' eigenvector of a less trivial type than here).

The model discussed here was discovered by the author in 1989~\cite{k1}.
Later on, a similar but different model was discovered
by J.~Hietarinta~\cite{h}, and then it was shown
by S.M.~Sergeev, V.V.~Mangazeev, Yu.G.~Stroganov~\cite{mss} that
both those models are particular cases of one model, parallel
in some sense to the Zamolodchikov model. This allows one to hope
that the reach mathematical structures already discovered in
connection with our model will be extended some time onto
the Zamolodchikov model as well.

\end{document}